\documentclass[aps,prc,superscriptaddress,amsfonts,amssymb,amsmath,bibnotes,footinbib,twocolumn,fleqn,10pt]{revtex4-1}

\usepackage{graphicx}
\usepackage{subfigure}
\usepackage{leftidx}
\usepackage{setspace}
\usepackage{array}
\usepackage{lineno,hyperref}
\usepackage{amssymb}
\usepackage{amsmath}

\begin{document}
\begin{spacing}{1.0}

\title{The correlation between quarter-point angle and nuclear radius}



\author{W.~H.~Ma}
\email[]{weihuma@impcas.ac.cn}
\affiliation{Key Laboratory of High Precision Nuclear Spectroscopy and Center for Nuclear Matter Science, Institute of Modern Physics, Chinese Academy of Science, Lanzhou 730000, People's Reublic of China}
\affiliation{University of Chinese Academy of Science, Beijing, 100049, People's Reublic of China}
\affiliation{Lanzhou University, Lanzhou 730000, China}
\author{J.~S.~Wang}
\email[]{jswang@impcas.ac.cn}
\affiliation{Key Laboratory of High Precision Nuclear Spectroscopy and Center for Nuclear Matter Science, Institute of Modern Physics, Chinese Academy of Science, Lanzhou 730000, People's Reublic of China}
\author{S.~Mukherjee}
\affiliation{Key Laboratory of High Precision Nuclear Spectroscopy and Center for Nuclear Matter Science, Institute of Modern Physics, Chinese Academy of Science, Lanzhou 730000, People's Reublic of China}
\affiliation{Physics Department, Faculty of Science, M.S. University of Baroda, Vadodara - 390002, India}
\author{Q.~Wang}
\affiliation{Key Laboratory of High Precision Nuclear Spectroscopy and Center for Nuclear Matter Science, Institute of Modern Physics, Chinese Academy of Science, Lanzhou 730000, People's Reublic of China}
\author{D.~Patel}
\affiliation{Physics Department, Faculty of Science, M.S. University of Baroda, Vadodara - 390002, India}
\author{Y.~Y.~Yang}
\affiliation{Key Laboratory of High Precision Nuclear Spectroscopy and Center for Nuclear Matter Science, Institute of Modern Physics, Chinese Academy of Science, Lanzhou 730000, People's Reublic of China}
\author{J.~B.~Ma}
\affiliation{Key Laboratory of High Precision Nuclear Spectroscopy and Center for Nuclear Matter Science, Institute of Modern Physics, Chinese Academy of Science, Lanzhou 730000, People's Reublic of China}
\author{P.~Ma}
\affiliation{Key Laboratory of High Precision Nuclear Spectroscopy and Center for Nuclear Matter Science, Institute of Modern Physics, Chinese Academy of Science, Lanzhou 730000, People's Reublic of China}
\author{S.~L.~Jin}
\affiliation{Key Laboratory of High Precision Nuclear Spectroscopy and Center for Nuclear Matter Science, Institute of Modern Physics, Chinese Academy of Science, Lanzhou 730000, People's Reublic of China}
\author{Z.~Bai}
\affiliation{Key Laboratory of High Precision Nuclear Spectroscopy and Center for Nuclear Matter Science, Institute of Modern Physics, Chinese Academy of Science, Lanzhou 730000, People's Reublic of China}
\author{X.~Q.~Liu}
\affiliation{Key Laboratory of High Precision Nuclear Spectroscopy and Center for Nuclear Matter Science, Institute of Modern Physics, Chinese Academy of Science, Lanzhou 730000, People's Reublic of China}



\date{\today}

\begin{abstract}
The correlation between quarter-point angle of elastic scattering and nuclear matter radius has been studied systematically. Various phenomenological formulae with parameters for nuclear radius are adopted and compared by fitting the experimental data of quarter point angle extracted from nuclear elastic scattering reaction systems. The parameterized formula related to binding energy is recommended, which gives a good reproduction of nuclear matter radii of halo nuclei. It indicates that the quarter-point angle of elastic scattering is quite sensitive to the nuclear matter radius and can be used to extract the nuclear matter radius.
\end{abstract}

\pacs{}

\maketitle

\section{Introduction}
In the recent years, the nuclear reactions with unstable/weakly bound nuclei that have low breakup threshold and exotic structure has shown remarkable features that is different from those of tightly bound nuclei. It will be interesting to understand and revisit in detail, the difference in the reaction mechanisms using a tightly, weakly and unbound or halo nuclei. Very recently, a phenomenological comparison of reduced reaction cross sections of different reaction systems was proposed by using $Wong^{'}s$ model \cite{bibitem1,bibitem2,bibitem3}. Several authors have extracted the quarter-point angle from the elastic scattering angular distribution reaction cross section, in order to compare the weakly and tightly bound projectiles \cite{bibitem4,bibitem5,bibitem6}. The quarter point angle which is also called the ¡°grazing angle¡± or ¡°rainbow angle¡± is one of the most conspicuous features of heavy-ion elastic scattering at above-barrier energies. Accordingly, the radius of interaction $R_{int}$ correlating with quarter-point angle, is the sum of projectile and target radius and approximately equals to the classical apsidal distance, the distance of closest approach, evaluated at the energy for which the experimental cross section is one-quarter of the corresponding Rutherford cross section \cite{bibitem7}. Earlier evaluatation of the $R_{int}$ is given by $r_{0}(A_{p}^{1/3} + A_{t}^{1/3})$, where $A_{p}$ and $A_{t}$ are the mass numbers of projectile and target, respectively. It has been found that the value of $r_{0}$ ranging from 1.20 to 1.30 fm are the most appropriate values for the heavy ion interaction at energies $\geq 10.0 $ MeV/u (Baluch et al., 1998 and references therein).

Experimentally, the nuclear radius (or nuclear matter distribution) can be determined by the measurement of electron scattering, isotope shift and interaction cross section etc. Since the electron is structureless and the electromagnetic interaction is very well known, therefore the charge distribution of a nucleus can be precisely obtained from the electron scattering measurements. The proton distribution can be deduced from the nuclear charge distribution in the case of stable nuclei. However, for an unstable nuclei being short-lived and difficult to use as a target, especially for the halo nuclei, one usually uses the isotope method or interaction cross section measurements to determine the size of nucleus. To be more precise, one can say that electron scattering is better than isotope shift and isotope shift is better than interaction cross section. Moreover the interaction cross section is much model dependent. Earlier, it was observed that  the nuclear size is obviously correlated to the quarter-point angle. This is because the quarter-point angle is a function of $R_{int}$. This indicates that we may extract the radius of unstable nuclei from the experimental quarter-point angle. This could be a new experimental method to determine the nuclear size.

As introduced above, interaction radius can be extracted from the quarter-point angle through the elastic scattering angular distribution of the reacting system. Based on the concept of quarter-point angle, the main objective of this work is to compare the tightly bound, weakly bound (stable) and halo projectiles using the phenomenological formula for interaction radius. This difference of the three kinds of projectiles can be employed to find a better understanding of interaction radius. Furthermore, nuclear radius can also be obtained from this analysis.

\section{Phenomenological formulae with parameters for radius of interaction}
The theoretical quarter-point angle described elsewhere \cite{bibitem6}, \begin{equation} \theta_{1/4} = 2\arcsin{[1/(2x-1)]}. \end{equation} Which is a function of the dimensionless variable $x$, where $x=E_{cm}/V_{coul}$, the ratio of the center of mass energy $E_{cm}$ to Coulomb barrier $V_{coul}$. The experimental values of the quarter-point angle were extracted from the available experimental data of the elastic scattering by fitting the angular distribution of the differential cross sections with the optical model. The corresponding center-of-mass energies can be also obtained from experiments. The Coulomb barrier was determined by $Z_{p}Z_{t}e^{2}/R_{int}$, where $Z_{p}$ and $Z_{t}$ are the number of protons in the projectile and target respectively. In addition, the value of $R_{int}$ can be obtained from the experimental values of quarter-point angle by the following relationship. $\frac{Z_{p}Z_{t}e^{2}}{2E_{cm}}\cdot{(1+\frac{\csc(\theta_{1/4})}{2}))}$. The experimental values of quarter point angle are given in APPENDIX.

The introduction of the reduced energy parameter $x$ is very useful to compare the quarter-point angle of different reaction systems together in one graph. FIG. 1 shows the comparison of quarter point angles obtained from a large amount of experimental data, using different tightness of projectiles, i.e., tightly bound, weakly bound, and halo projectiles. In general, the experimental values of quarter point angle of these three different type of projectiles successively decrease when the value of $x$ is fixed. However, all the experimental points of quarter-point angle are obviously lower than the curve of the theoretical quarter-point angle function (TQAF). This is because the theoretical $R_{int}$ is simply given by phenomenological formula $A_{p}^{1/3} + A_{t}^{1/3}$ ($PF_{1}$, as employed by L. Jin et al. \cite{bibitem6} to calculate the value of $x$.

Additionally, from FIG. 1, it can be observed that the comparison of the three kinds of projectiles is less distinguishable when a large amount of experimental data is taken into account, although the values
of the quarter point angle follow a successively decreasing general trend from the tightly bound to the halo for a fixed value of $x$. In general, for tightly bound systems for a given value of $x$ from 0.8 to 2.0, experimental points show larger underestimation as compared to the higher values of $x$. This anomaly was observed while deriving $R_{int}$ by using the usual formula ($A_{p}^{1/3} + A_{t}^{1/3}$). Similar wide distribution can also be observed in the case of weakly bound and halo systems.

\begin{figure}
\centering
\includegraphics[width=0.46\textwidth]{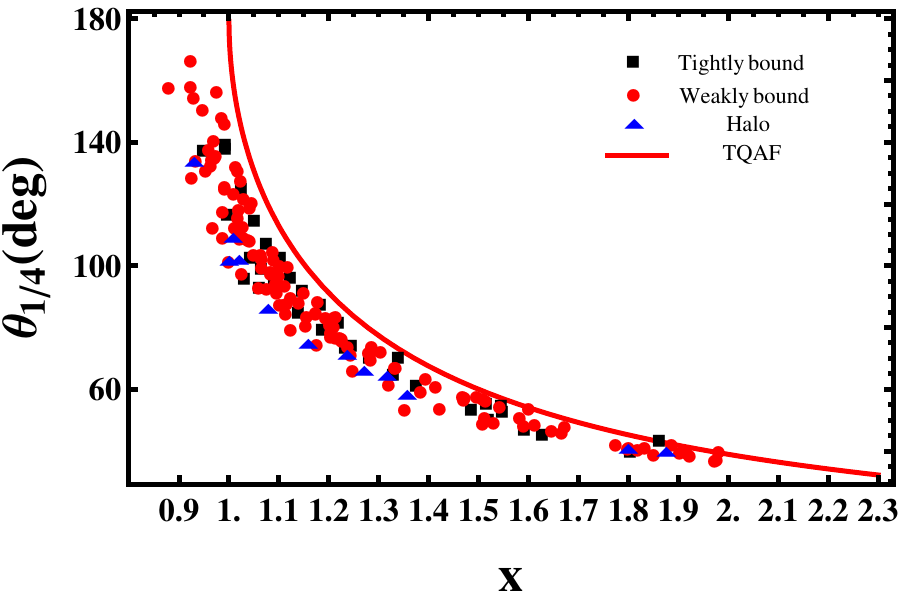}
 \caption{\label{Fig_1} (Color online) The quarter point angle as a function of reduced energy x in the interval from 0.8 to 2.0 for phenomenological formula $R_{int}=A_{p}^{1/3} + A_{t}^{1/3}$ ($PF_{1}$). The color points stand for the experimental quarter-point angles. The theoretical quarter-point angle function is labeled as TQAF.}
\end{figure}

According to above discussion, it is necessary to introduce a modified formula for $R_{int}$ as given by $a\cdot{A_{p}^{1/3}} + b\cdot{A_{t}^{1/3}}$ ($PF_{2}$), where the parameters $a$ and $b$ are fitted by extracting experimental $R_{int}$ separately for tightly bound, weakly bound and halo projectiles.

\begin{table}
\centering
\begin{tabular}{cccccccc}
\hline
Projectile & $a (fm)$  & $b (fm)$  \\
\hline
Tightly bound & $1.123$ & $1.00$  \\
Weakly bound & $1.187$ & $1.00$  \\
Halo & $1.333$ & $1.00$  \\
\hline
\end{tabular}
\caption{The fitted values of $a$ and $b$ for the tightly bound, weakly bound and halo projectiles.}
\end{table}

In the fitted values, as shown in Table I, the parameter b (b = 1) is kept constant for the all three kind of systems assuming the target to be stable. The fitted parameter $a$ increases from tightly bound to halo projectiles. It can be distinctly observed that the fitted values of parameter $a$ for tightly bound, weakly bound and halo projectiles describe the difference between the reaction systems and well indicate the size of projectiles from the expression $a\cdot{A_{p}^{1/3}}$. This is in accordance with the nuclear size obtained from previous studies. This also clearly indicates that the radius calculated by the expression $A_{p}^{1/3} + A_{t}^{1/3}$ in order to compare the different kinds of projectiles result in the underestimation of the size of weakly bound projectiles, and even more in the case of halo projectiles.

Thus the modified expression of $R_{int}$ can reduce the deviation between the experimental data and the theoretical curve of the quarter-point angle as a function of $x$. In FIG. 2 , the result of the modification is shown. When compared with the points in FIG. 1, the deviation between the experimental data and the theoretical curve and among the three kinds of projectiles is diminished.

\begin{figure}
\centering
\includegraphics[width=0.46\textwidth]{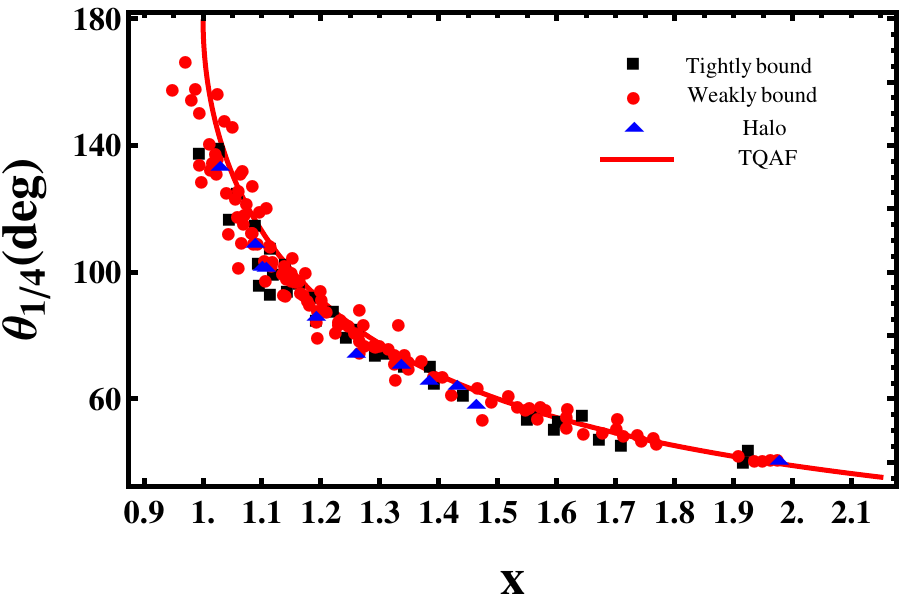}
 \caption{\label{Fig_2} (Color online) Same as FIG. 1,  except for phenomenological formula ($PF_{2}$).}
\end{figure}

With the enlightenment of reducing the difference in the quarter point angle values among the three kinds of projectiles by using modified $R_{int}$, it is feasible to find a better phenomenological formula to give a better description of the strong absorption radius. To begin with, the nuclear radius based on the liquid drop model was used. Additionally, nucleon distribution having finite surface thickness was assumed instead of uniform distribution. Considering the isospin symmetry that distinguishes between proton and neutron, affected equally by the nuclear strong force, the improved expression of $R_{int}$ with parameters $r_{0}$, $r_{1}$, $r_{2}$ and $a_{z}$ is given by ($PF_{3}$) \cite{bibitem8},
\begin{equation} \begin{aligned} R_{int} &=& \sum_{i=p,t}R_{i} ;  \end{aligned} \end{equation}
\begin{equation} \begin{aligned}
R_{i}=((r_{0}+\frac{r_{1}}{A_{i}^{2/3}}+\frac{r_{2}}{A_{i}^{4/3}})
+a_{z}\frac{Z_{i}-Z_{stable}}{A_{i}})A_{i}^{1/3}; \end{aligned} \end{equation}
where $Z_{stable}=\frac{A_{i}}{1.98+0.016\cdot{A_{i}^{2/3}}}$; $i$ denotes projectile and target. The values of the parameters $r_{0}$, $r_{1}$, $r_{2}$ and $a_{z}$ obtained by fitting the experimental data, were $1.0152 fm$, $0.6383 fm$, $-1.2781 fm$  and $-0.2981 fm$  respectively. However, as shown in FIG. 3 (a), the quarter point angle values are still inconsistent with respect to the three types of projectiles, although the deviation between the experimental data and the theoretical curve show a decrease. Thus the modified formula given by $PF_{3}$ is not adequate to accurately describe the nuclear size, especially for the halo nuclei. In order to obtain a consistent description for nuclear size, it is important to discuss this phenomenological formulation with further improvement by considering the binding energy of the nuclei. According to the quantum mechanics, the nuclear rms radius is inversely proportional to the binding energy. The relation between the rms radii and the binding energy can be obtained from the simplified N single-particle Schr\"{o}dinger equation as, $R(B)=\frac{4.04}{\sqrt{B(A)}} $\cite{bibitem9}. However, this relation is not sufficient for a good agreement with the experimental data. A modified quantitative formula of the nuclear rms radius as a function of binding energy per nucleon was introduced and discussed by Wang et al., \cite{bibitem10}. It is important to take into account the binding energies and based on this fact, the theoretical $R_{int}$ is given by ($PF_{4}$)
\begin{equation} \begin{aligned} R_{int} &=& \sum_{i=p,t}R_{i} ;  \end{aligned} \end{equation}
\begin{equation} \begin{aligned}
R_{i}&=&\lambda_{0}A_{i}^{1/3}+\lambda_{1}+\lambda_{2}\frac{I_{i}}{\sqrt{B_{i}}}+(\lambda_{3}\frac{I_{i}}{\sqrt{B_{i}}})^{2}; \end{aligned} \end{equation}
where $ I_{i}=\frac{A_{i}-2Z_{i}}{A_{i}}$, denotes the symmetry parameter, $ B_{i}$ is the binding energy per nucleon and subscript denotes projectile and target. The experimental data of quarter-point angle were fitted to obtain the parameters $\lambda_{0}=0.9776 fm$, $\lambda_{1}=0.2475 fm$, $\lambda_{2}=-0.1492 fm/MeV^{1/2}$ and $\lambda_{3}=10.7186 fm/MeV$. The experimental values of binding energy were taken from literature \cite{bibitem11} and were used to determine $B_{i}$. As a result, the deviation between the experimental and the theoretical curve for all the three types of projectiles (mainly between the tight bound and the halo) were diminished as can be seen in FIG. 3 (b). The goodness of fit for $PF_{3}$ and $PF_{4}$ was obtained by square of regression fit (R-Square). R-Square is a number that indicates how well data fit a curve. An R-Square of 1 indicates that the regression line perfectly fitting the data, while an R-Square of 0 the line does not fit the data at all. In the present work R-Square for $PF_{3}$ is 0.998585 and for $PF_{4}$ 0.9986632. So $PF_{4}$ shows better fit than $PF_{3}$. Using the results from the above method and comparing them with the fitted data as shown in FIG. 2, one can emphasize the importance of the binding energy to understand the nuclear size, especially in the case of halo nucleus. For a clear justification, the next section gives a more detailed comparison.
\begin{figure}
\centering
\includegraphics[width=0.46\textwidth]{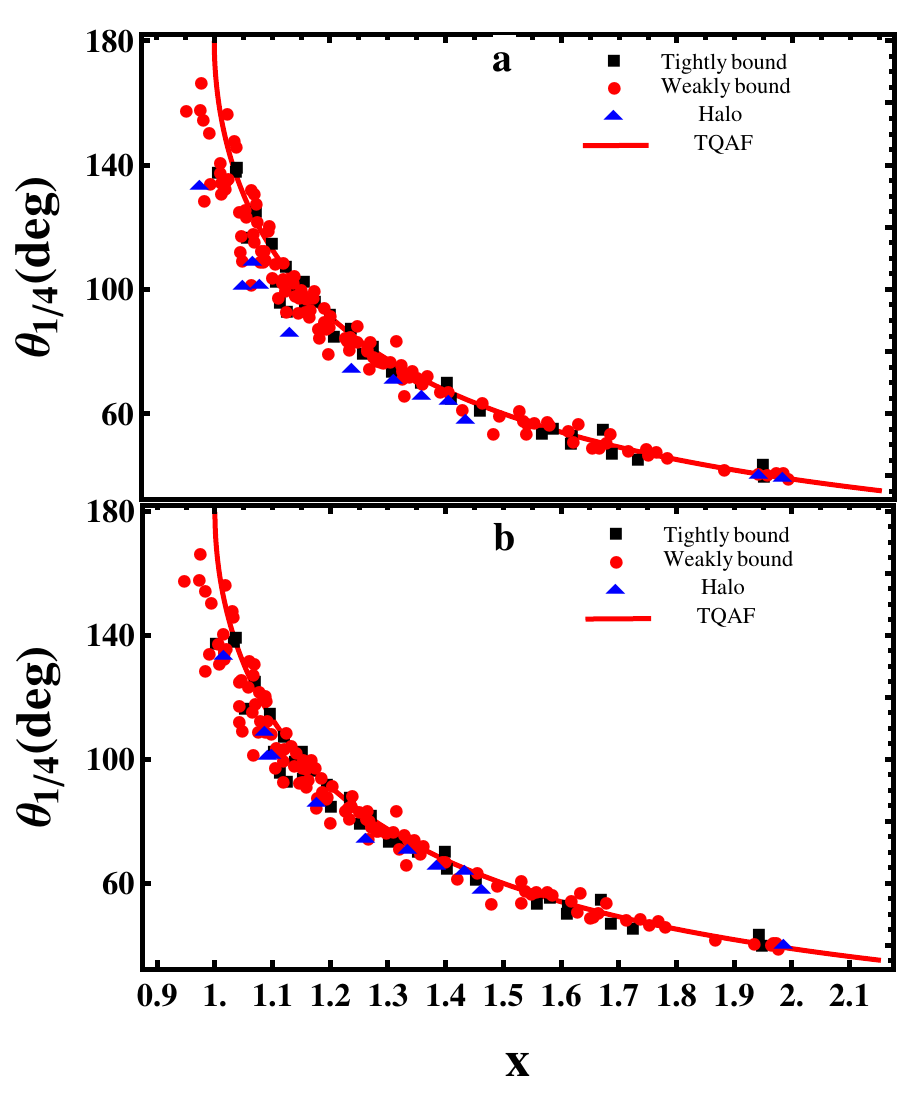}
 \caption{\label{Fig_3} (Color online) Theoretical and experimental Quarter point angle values for three types of projectiles using $PF_{3}$ (a) and $PF_{4}$ (b). The color points are for the experimental quarter-point angle values. The experimental data were used in the interval of x from 0.8 to 2.0. The theoretical quarter-point angle function is labeled as TQAF.}
\end{figure}

\section{Comparison of the four phenomenological formulae}
In this section, we give a quantitative comparison of the four phenomenological formulae ($PF_{1}$ to $PF_{4}$). We define a goodness of fit ratio $\eta=\frac{x-x_{PF}}{x_{PF}}$ that will estimate the deviation between the experimental quarter-point angle and the curve of (TQAF). In other words, smaller is the value of $\eta$, less will be the deviation. In this relation $x_{PF}$ is determined by the center-mass energy and the Coulomb barrier via the phenomenological formulae, $x_{PF}=E_{cm}/V_{coul}$. However, $x$ is directly calculated by using the formula (1), namely, using the curve of TQAF. Both $x_{PF}$ and $x$ corresponds to the same quarter-point angle extracted from elastic scattering angular distribution of the reacting systems. The comparison of $\eta$ in terms of the four phenomenological formulae ($PF_{1}$ to $PF_{4}$) is shown in FIG. 4.  It is clear from the figure that the modified formulae $PF_{2}$, $PF_{3}$ and $PF_{4}$ show a better agreement as compared to $PF_{1}$ so far as the deviation between the experimental data and the theoretical curve (TQAF) is concerned. The advantages and disadvantages of the four methods for calculating $R_{int}$ are more clearly understood while comparing the deviation between tightly bound and halo projectiles. This deviation between tightly bound and halo projectiles by the four methods may be more clearly obtained by introducing another parameter $\triangle{\eta}=\overline{\eta_{H}}-\overline{\eta_{TB}}$, where, $\overline{\eta_{H}}$ is the arithmetic mean of $\eta$ for halo nuclei and $\overline{\eta_{TB}}$ is that for tightly bound nuclei. In addition, the inconsistency of the experimental points can be found for tightly bound systems with two clear groups (taking FIG. 5 as a sample case) while for the weakly bound systems there is a scattered distribution. We take the case of tightly bound nuclei to explain this feature. Analogically, the exact deviation between $\overline{\eta_{1-TB}}$ and $\overline{\eta_{2-TB}}$ is given by $\triangle{\eta_{TB}}=\overline{\eta_{1-TB}}-\overline{\eta_{2-TB}}$, where $\overline{\eta_{1-TB}}$ is the arithmetic mean of  for the group of experimental quarter-point angle points of tight bound nuclei far from the curve of TQAF and $\overline{\eta_{2-TB}}$ is that near to the curve of TQAF. The calculated values of $\triangle{\eta}$ and $\triangle{\eta_{TB}}$ are shown in TABLE II. From these values, it may be observed that the results given by $PF_{4}$ show the minimum values for both $\triangle{\eta}$ and $\triangle{\eta_{TB}}$. Therefore, one can conclude that among all the four phenomenological formulae, $PF_{4}$ shows the best improvement in present work for all the three types of nuclei. Thus in the present work, $PF_{4}$ is recommended for giving better consistency among tightly bound systems, besides reducing the deviation between the experimental data and the theoretical curve of quarter-point angle.
\begin{table}
\centering
\begin{tabular}{cccccccc}
\hline
 & $PF_{1}$  & $PF_{2}$ & $PF_{3}$ & $PF_{4}$  \\
\hline
$\triangle{\eta}$ & $0.0374$ & $0.0077$ & $0.0342$ & $0.0059$  \\
$\triangle{\eta_{TB}}$ & $0.056$ & $0.040$ & $0.039$ & $0.033$  \\
\hline
\end{tabular}
\caption{The values of $\triangle{\eta}$ and $\triangle{\eta_{TB}}$ for $PF_{1}$, $PF_{2}$, $PF_{3}$ and $PF_{4}$. }
\end{table}

\begin{figure}
\centering
\includegraphics[width=0.46\textwidth]{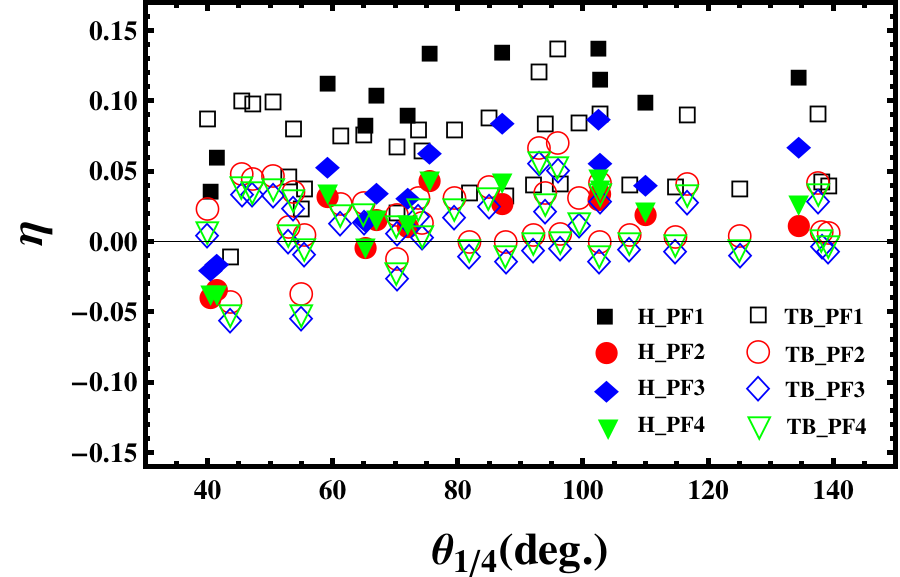}
 \caption{\label{Fig_4} (Color online) Comparison of the four phenomenological formulae with the experimental data used in the interval of x from 0.8 to 2.0. The solid symbols are for the halo nuclei and the open symbols for tightly bound nuclei.}
\end{figure}

\begin{figure}
\centering
\includegraphics[width=0.46\textwidth]{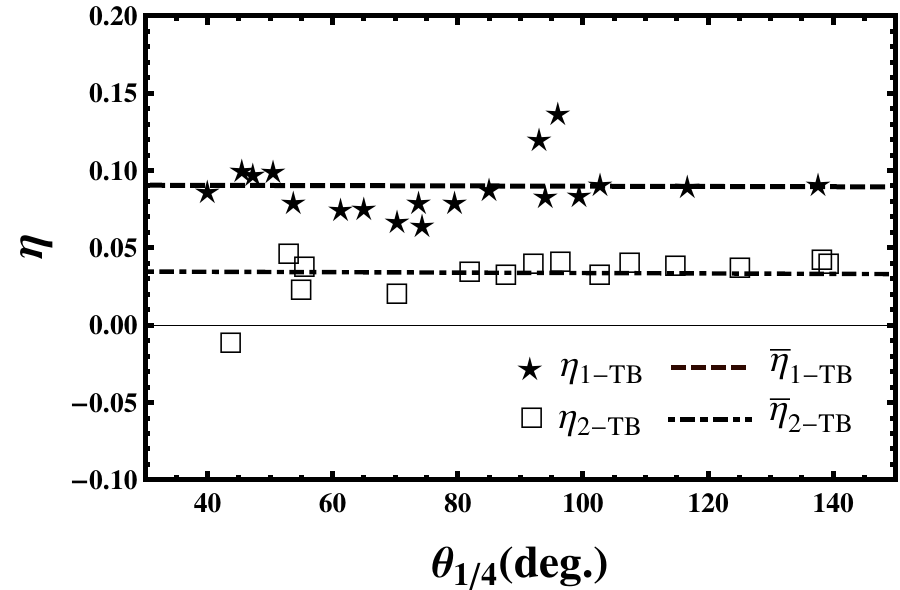}
 \caption{\label{Fig_5} (Color online) Comparison of $\eta$ with the experimental data used in the interval of x from 0.8 to 2.0 for the sets of experimental quarter-point angle points of the tight bound nuclei near and far from the curve of TQAF for $PF_{1}$. $\eta_{1-TB}$ is the points of $\eta$ for that far from the curve of TQAF and $\eta_{2-TB}$ is the points of $\eta$ for that near the curve of TQAF.}
\end{figure}

In conclusion, the theoretical radius of interaction given by $PF_{1}$ is not the good formulae although it can be used to compare the different tightness of systems. The $PF_{2}$ with parameters separately fitted by tight bound, weakly bound and halo projectiles is a parameterized method to do the comparison. For finding the unified formula for nuclear radius, namely for the radius of interaction, the $PF_{3}$ and $PF_{4}$ are compared. However, the improvement in $PF_{3}$ with symmetry dependence considered is not good enough to consistently describe the nuclear size. Finally, the $PF_{4}$ with the binding energy considered is recommended by this work for the phenomenological formula of $R_{int}$, which reduces the deviation not only between the experimental data and the theoretical curve but also among the three kinds of projectiles. And the inconsistency among the tightly bound systems is also improved by $PF_{4}$.
\section{Calculation of Nuclear Radius}
As discussed above, $PF_{4}$ not only reduces the deviation between the experimental data and the theoretical curve but also among the three kinds of projectiles, when we compare the four parameterized theoretical formula ($PF_{1}$ to $PF_{4}$) in order to obtain the radius of interaction $R_{int}$. Which means that if we use the recommended formula $PF_{4}$ for calculating $R_{int}$ of systems with tightly bound, weakly bound and halo projectiles to calculate $x$ and plot the figure of $\theta_{1/4}$ vs $x$, we can see that all the systems are on one curve of TQAF. The nuclear size is obviously correlated to the quarter-point angle, because $x$ is a function of $R_{int}$, which indicates that we may extract the radius of nuclei from the experimental quarter-point angle.

\begin{table}
\centering
\begin{tabular}{cccccccc}
\hline
Nuclei & $R_{rms}/fm$  & Ref. \\
\hline
${}^{6}He$ & $2.30\pm{0.07}$ & \cite{bibitem15}  \\
${}^{8}He$ & $2.69\pm{0.03}$ & \cite{bibitem16}  \\
${}^{8}B$ & $2.38\pm{0.04}$ & \cite{bibitem17}  \\
${}^{9}C$ & $2.71\pm{0.32}$ & \cite{bibitem18}  \\
${}^{10}Be$ & $2.479\pm{0.028}$ & \cite{bibitem19}  \\
${}^{10}C$ & $2.42\pm{0.10}$ & \cite{bibitem20}  \\
${}^{11}Li$ & $3.34_{-0.08}^{+0.04}$ & \cite{bibitem21}  \\
${}^{11}Be$ & $2.73\pm{0.05}$ & \cite{bibitem22}  \\
${}^{11}C$ & $2.46\pm{0.03}$ & \cite{bibitem19}  \\
${}^{12}N$ & $2.47\pm{0.07}$ & \cite{bibitem23}  \\
${}^{13}O$ & $2.53\pm{0.05}$ & \cite{bibitem23}  \\
${}^{14}Be$ & $3.10\pm{0.15}$ & \cite{bibitem24}  \\
${}^{17}B$ & $2.99\pm{0.09}$ & \cite{bibitem24}  \\
${}^{17}F$ & $2.71\pm{0.18}$ & \cite{bibitem25} \\
${}^{17}Ne$ & $2.75\pm{0.07}$ & \cite{bibitem23}  \\
${}^{19}B$ & $3.11\pm{0.13}$ & \cite{bibitem24}  \\
${}^{23}Al$ & $2.905\pm{0.250}$ & \cite{bibitem25}  \\
${}^{27}P$ & $3.020\pm{0.155}$ & \cite{bibitem25}  \\
\hline
\end{tabular}
\caption{The experimental values of rms matter radii for light halo nuclei.}
\end{table}

For all applications, we calculate radius of nuclei as projectiles using the formulae fitted from experimental quarter-point angle and separately compare them with the experimental charge radius \cite{bibitem12} for tightly-bound (FIG. 6) and weakly-bound nuclei (FIG. 7) and with experimental root mean square (rms) matter radius (TABLE III) for halo nuclei (FIG. 8). For light nuclei, the nuclear experimental charge radius usually agrees with the mass radius, but for heavy nuclei having more neutrons than protons, the mass radius might larger than the charge radius. As shown in Figs. 7 and 8, the calculation of tightly-bound and light weakly-bound nuclei based on $PF_{3}$ and $PF_{4}$ can be deemed as a better representation for determining nuclear size than $PF_{1}$ and $PF_{2}$. The goal of extracting the radius of exotic nuclei from the experimental quarter-point angle can be embodied by the calculation based on $PF_{4}$ as shown in FIG. 8, which clearly shows the feasibility of acquiring the radius of halo nuclei via the experimental quarter-point angle. From the comparison, the calculation for light halo nuclei based on $PF_{3}$ does not agree with the experimental values.

\begin{figure}
\centering
\includegraphics[width=0.48\textwidth]{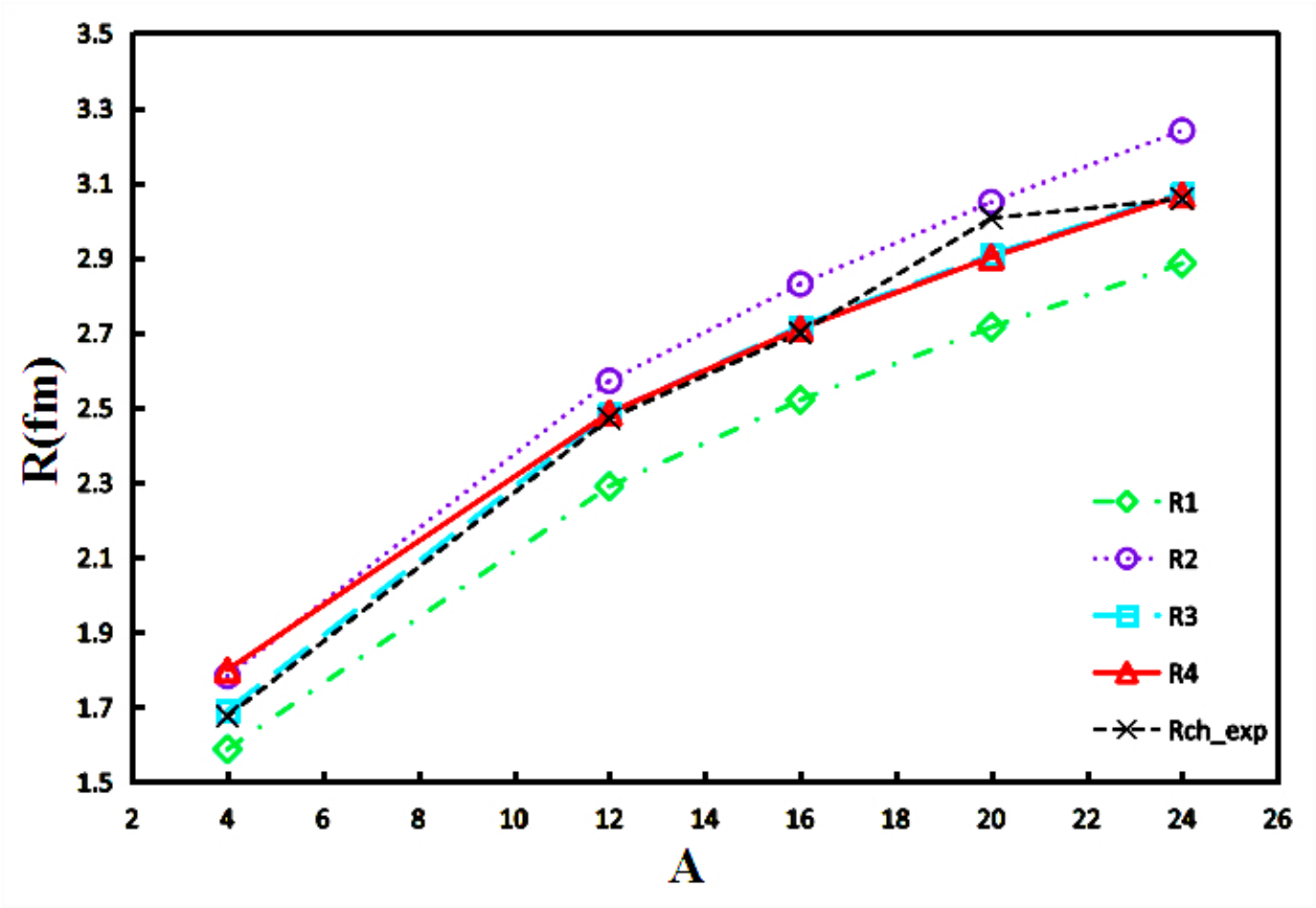}
 \caption{\label{Fig_6} (Color online) Comparison of the calculated mass radius (R1, R2, R3 and R4) basing on the four phenomenological formulae with nuclear charge radius for tightly-bound nuclei ${}^{4}He$, ${}^{12}C$, ${}^{16}O$, ${}^{20}Ne$ and ${}^{24}Mg$.}
\end{figure}
\begin{figure}
\centering
\includegraphics[width=0.48\textwidth]{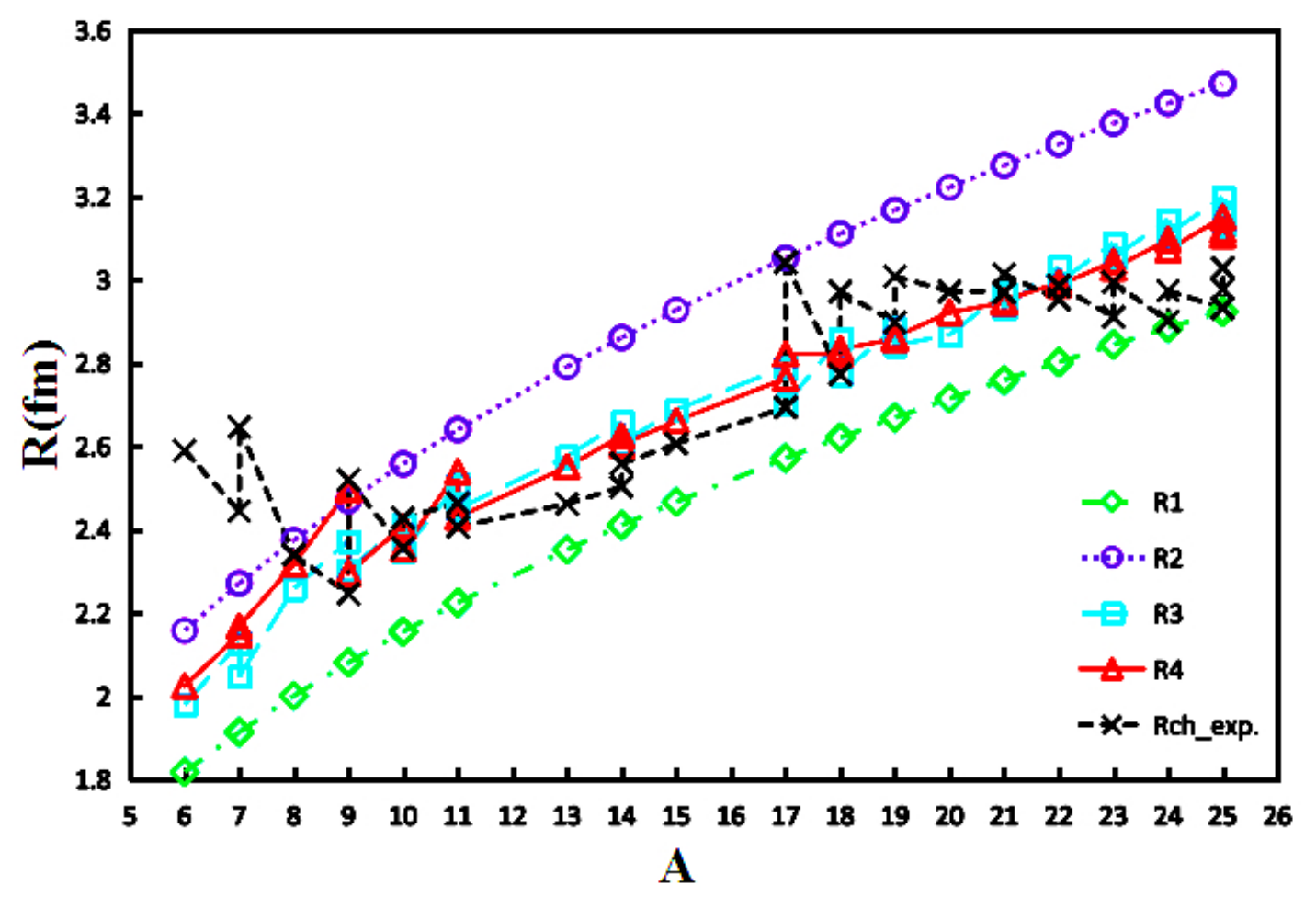}
 \caption{\label{Fig_7} (Color online) Comparison of the calculated mass radius based on the four phenomenological formulae $PF_{1}$ to $PF_{4}$ (R1, R2, R3 and R4) with nuclear rms charge radius for weakly-bound nuclei.}
\end{figure}
\begin{figure}
\centering
\includegraphics[width=0.49\textwidth]{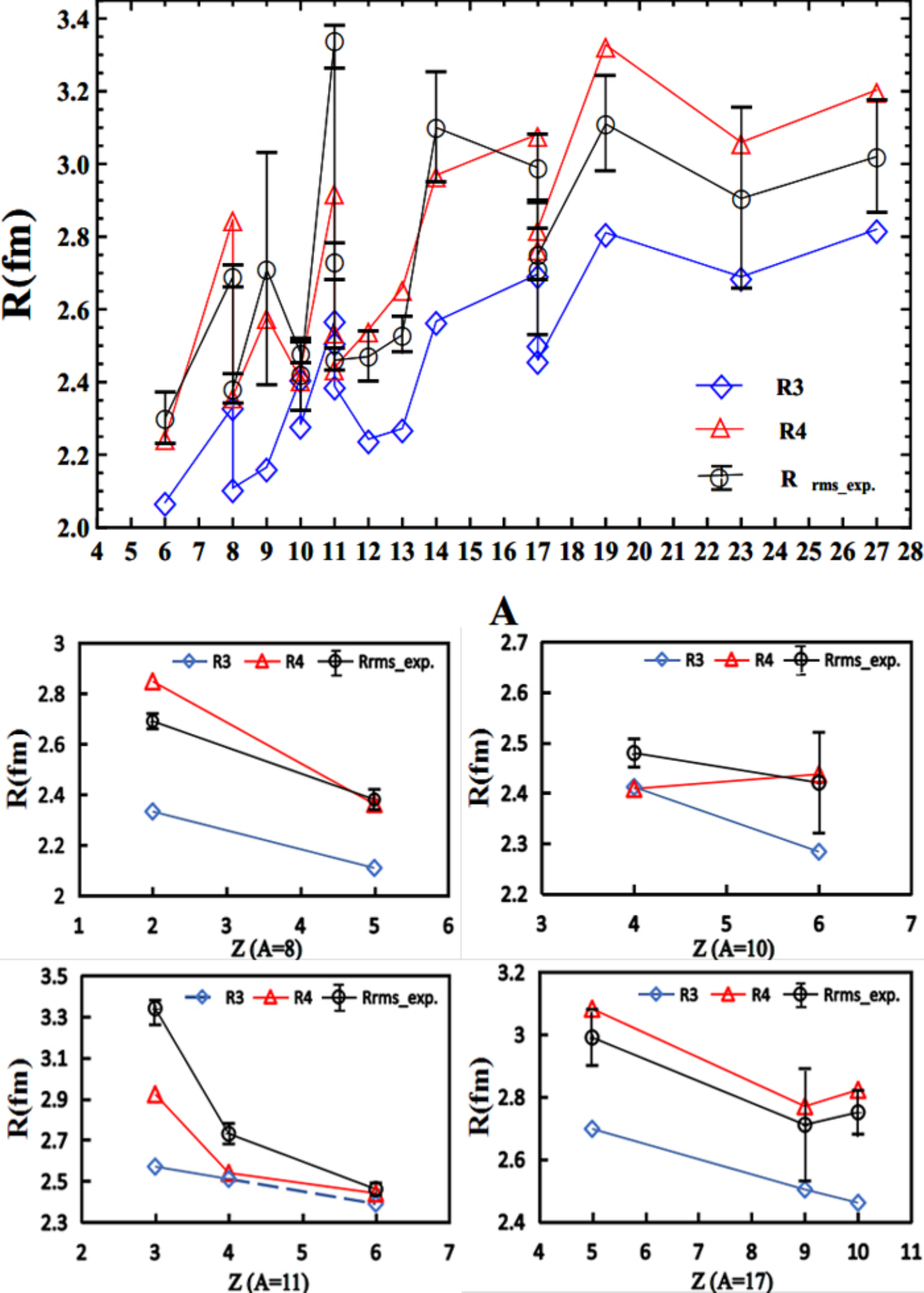}
 \caption{\label{Fig_8} (Color online) Comparison of the calculated mass radius based on the phenomenological formulae $PF_{3}$ and $PF_{4}$ (R3 and R4) with nuclear rms matter radius for halo nuclei.}
\end{figure}

As a matter of fact, the elastic scattering of a halo nucleus from a stable target can give simple direct evidence for the structure of the halo nucleus \cite{bibitem13}. The angular distribution of elastic scattering reactions show a maximum difference for incident energies around the top of the Coulomb barrier, thereby suggesting that it is in this energy region where the elastic scattering is most sensitive to the details of the nuclear structure of the exotic projectiles \cite{bibitem14}. The quarter-point angle as a function of radius of interaction $R_{int}$ obtained via angular distributions of elastic scattering cross section, is related to the actual reaction mechanisms, which is not only related to the size of the nuclei but also to the elastic scattering reactions with coupling to the channels of inelastic scattering or other reactions. The nuclear properties, such as, nuclear radius, isospin symmetry and binding energy per nucleon, will affect the strength of the couplings for different incident energies. Inversely, we can extract the information of the structure from elastic scattering reaction, such as the nuclear size. Experimentally, the nuclear radius can be determined by electron scattering, isotope shift and interaction cross section etc. Since electron is structure less and the electromagnetic interaction is known very well, therefore the charge distribution of the nuclei can be precisely measured by electron scattering. However, it is suitable for the stable nuclei only, as the unstable nuclei are short-lived and difficult to use as targets. For the unstable nuclei especially for the halo nuclei, people usually use the isotope method or interaction cross section to measure the size of nuclei. However, extracting the radius of unstable nuclei from the experimental quarter-point angle could be a useful tool as the new experimental measurement. Therefore, the $PF_{4}$ with more details of structure (spatial extension, isospin symmetry and binding energy) is recommended by this work for nuclear radius.
\section{Conclusions}
The motivation of present work is to correlate quarter-point angle and nuclear radius. The theoretical radius of interaction $R_{int}$ were obtained and compared. In this work, four phenomenological formulae ($PF_{1}$ to $PF_{4}$) were presumed and the parameters for different formulae were fitted by using the extracted experimental values of $R_{int}$. Considering the different kinds of reaction systems, the four phenomenological formulae are analyzed and discussed. Based on the above mentioned formulae, the radii of different kind of nuclei as projectiles were obtained and explained in detail. As a result, the parameterized formula related to binding energy was recommended. In conclusion, the deviation between the experimental data and the theoretical curve and among the three kinds of projectiles can be minimized by appropriately calculating the nuclear radius in order to determine the radius of interaction. This may lead to a better understanding of the nuclear structure and the actual reaction mechanisms using the three types of projectiles (strongly bound, weakly bound and halo nuclei).
\section{Acknowledgments}
This work was financially supported by the National Natural Science Foundation of China with Grant No. U1432247 and 11575256 and the National Basic Research Program of China (973 Program) with Grant No. 2014CB845405 and 2013CB83440x. One of the authors (SM) thanks the Chinese Academy of Sciences for the support in the form of President¡¯s International Fellowship Initiative (PIFI) Grant No. 2015-FX-04.

\section*[level (1)]{Appendix}
The selected experimental quarter-point angle data for tightly bound and halo systems are arranged in the following table. This data was obtained from literature in [http://nrv.jinr.ru/nrv/]. The data for weakly bound projectiles, ${}^{6}Li$,${}^{7}Li$,${}^{9}Be$,${}^{11}B$,${}^{14}N$,${}^{15}N$ and ${}^{19}F$ can be also found in [http://nrv.jinr.ru/nrv/].

\begin{table}
\centering
\begin{tabular}{cccccccc}
\hline
Reaction Systems &  $  E_{cm}(MeV) $   &  $ \theta_{cm}(deg.) $  & Reaction Systems & $  E_{cm}(MeV)  $  & $ \theta_{cm}(deg.) $ \\
\hline
${}^{12}C+{}^{28}Si$ & $16.80$ & $99.4$ &
${}^{12}C+{}^{28}Si$ & $34.51$ & $31.0$\\
${}^{12}C+{}^{28}Si$ & $45.50$ & $24.8$ &
${}^{12}C+{}^{28}Si$ & $130.48$ & $6.4$\\
${}^{12}C+{}^{208}Pb$ & $59.47$ & $139.3$ &
${}^{12}C+{}^{208}Pb$ & $61.36$ & $125.1$\\
${}^{12}C+{}^{208}Pb$ & $66.09$ & $102.7$ &
${}^{12}C+{}^{208}Pb$ & $70.81$ & $87.7$\\
${}^{12}C+{}^{208}Pb$ & $80.27$ & $70.3$ &
${}^{12}C+{}^{208}Pb$ & $90.76$ & $55.5$\\
${}^{12}C+{}^{208}Pb$ & $111.57$ & $43.7$ &
${}^{12}C+{}^{208}Pb$ & $170.18$ & $24.6$\\
${}^{12}C+{}^{208}Pb$ & $283.64$ & $14.0$ &
${}^{12}C+{}^{208}Pb$ & $397.09$ & $9.6$\\
${}^{16}O+{}^{12}C$ & $10.29$ & $96.0$ &
${}^{16}O+{}^{12}C$ & $15.43$ & $55.0$\\
${}^{16}O+{}^{12}C$ & $18.00$ & $40.0$ &
${}^{16}O+{}^{12}C$ & $26.57$ & $30.0$\\
${}^{16}O+{}^{12}C$ & $34.29$ & $20.8$ &
${}^{16}O+{}^{12}C$ & $56.57$ & $12.3$\\
${}^{16}O+{}^{16}O$ & $37.50$ & $22.7$ &
${}^{16}O+{}^{16}O$ & $40.50$ & $20.6$\\
${}^{16}O+{}^{16}O$ & $43.50$ & $19.5$ &
${}^{16}O+{}^{16}O$ & $46.00$ & $18.0$\\
${}^{16}O+{}^{16}O$ & $47.50$ & $17.3$ &
${}^{16}O+{}^{16}O$ & $58.00$ & $13.6$\\
${}^{16}O+{}^{40}Ca$ & $28.57$ & $93.0$ &
${}^{16}O+{}^{40}Ca$ & $33.57$ & $74.3$\\
${}^{16}O+{}^{40}Ca$ & $42.86$ & $47.3$ &
${}^{16}O+{}^{40}Ca$ & $40.55$ & $50.5$\\
${}^{16}O+{}^{40}Ca$ & $43.45$ & $45.5$ &
${}^{16}O+{}^{56}Fe$ & $31.11$ & $137.6$\\
${}^{16}O+{}^{56}Fe$ & $32.67$ & $116.7$ &
${}^{16}O+{}^{56}Fe$ & $34.22$ & $102.8$\\
${}^{16}O+{}^{56}Fe$ & $35.78$ & $94.0$ &
${}^{16}O+{}^{56}Fe$ & $37.33$ & $85.0$\\
${}^{16}O+{}^{56}Fe$ & $38.89$ & $79.5$ &
${}^{16}O+{}^{56}Fe$ & $40.44$ & $73.8$\\
${}^{16}O+{}^{56}Fe$ & $42.00$ & $70.3$ &
${}^{16}O+{}^{56}Fe$ & $43.56$ & $65.0$\\
${}^{16}O+{}^{56}Fe$ & $45.11$ & $61.3$ &
${}^{16}O+{}^{90}Zr$ & $67.92$ & $53.8$\\
${}^{16}O+{}^{90}Zr$ & $117.34$ & $26.4$ &
${}^{16}O+{}^{90}Zr$ & $165.06$ & $17.8$\\
${}^{16}O+{}^{208}Pb$ & $120.25$ & $53.0$ &
${}^{16}O+{}^{208}Pb$ & $178.29$ & $31.2$\\
${}^{16}O+{}^{208}Pb$ & $77.07$ & $138.2$ &
${}^{16}O+{}^{208}Pb$ & $81.71$ & $114.8$\\
${}^{16}O+{}^{208}Pb$ & $83.57$ & $107.5$ &
${}^{16}O+{}^{208}Pb$ & $87.29$ & $96.4$\\
${}^{16}O+{}^{208}Pb$ & $89.14$ & $92.1$ &
${}^{16}O+{}^{208}Pb$ & $94.71$ & $81.9$\\
${}^{6}He+{}^{65}Cu$ & $17.90$ & $41.5$ &
${}^{6}He+{}^{65}Cu$ & $27.50$ & $25.0$\\
${}^{6}He+{}^{120}Sn$ & $17.19$ & $75.5$ &
${}^{6}He+{}^{120}Sn$ & $18.86$ & $67.0$\\
${}^{6}He+{}^{120}Sn$ & $19.52$ & $65.3$ &
${}^{6}He+{}^{197}Au$ & $38.82$ & $40.5$\\
${}^{6}He+{}^{208}Pb$ & $21.38$ & $110.0$ &
${}^{6}He+{}^{208}Pb$ & $26.24$ & $72.0$\\
${}^{6}He+{}^{208}Pb$ & $28.77$ & $59.2$ &
${}^{6}He+{}^{208}Pb$& $53.46$ & $27.8$\\
${}^{6}He+{}^{209}Pb$ & $21.87$ & $102.8$ &
${}^{8}B+{}^{58}Ni$ & $22.23$ & $134.5$\\
${}^{8}B+{}^{58}Ni$ & $23.90$ & $102.5$ &
${}^{8}B+{}^{58}Ni$ & $25.75$ & $87.1$\\
\hline
\end{tabular}
\end{table}

\end{spacing}
\end{document}